\documentclass[journal=jacsat,manuscript=article]{achemso}
\usepackage[version=3]{mhchem} 
\usepackage[T1]{fontenc}       
\usepackage{graphicx}
\usepackage{dcolumn}
\usepackage{bm}
\usepackage{mathtools}
\usepackage{hyperref}
\usepackage{color}
\usepackage[a4paper]{anysize}
\def\gsim{\lower.35em\hbox{$\stackrel{\textstyle>}{\textstyle\sim}$}}
\def\lsim{\lower.35em\hbox{$\stackrel{\textstyle<}{\textstyle\sim}$}}

\author{Luis Brey}
\affiliation{Instituto de Ciencia de Materiales de Madrid, CSIC, 28049 Cantoblanco, Spain}
\email{brey@icmm.csic.es}

\title{Magnetic Skyrmionic Polarons}
\begin{document}
\begin{abstract}
We study a two-dimensional electron gas exchanged-coupled to a system of classical magnetic ions. For large Rashba spin-orbit coupling
a single electron can  become self-trapped in a  skyrmion spin texture self-induced in the magnetic ions system. This new quasiparticle 
carries electrical and topological charge as well  as a large spin, and we named it as magnetic skyrmionic polaron. 
We study the range of parameters; temperature, exchange coupling, Rashba coupling and  magnetic field, for which the magnetic skyrmionic polaron is  the fundamental state in the system. 
The dynamics of this quasiparticle is studied using the collective coordinate approximation, and we obtain that in presence of an electric field the new quasiparticle shows, because the chirality of the skyrmion, a Hall effect. Finally we argue that the magnetic skyrmionic polarons can be found in large Rashba spin-orbit coupling semiconductors as GeMnTe.

\end{abstract}
{\bf Keywords:} {Magnetic Skyrmions, Polarons, Magnetic Semiconductors}
\maketitle
\newpage
\par 
\noindent
In 1962  Skyrme introduced a model where particles as photons and neutrons show up as topological defects of  fields of mesons\cite{Skyrme:1962aa}. These particles were named skyrmions and also appear as topological excitations in the continuum 
limit of the two-dimensional (2D) ferromagnetic Heisenberg model;  the celebrated non-linear sigma model (NLSM)\cite{Belavin-1984}, 
$H_{NLSM}=\frac {\rho_s} 2 \int \partial _{\alpha} {\bf n} \cdot \partial _{\alpha} {\bf n} \, {\rm d}^2 {\bf r}$. Here ${\bf n}$ is the magnetization unit vector and $\rho _s$ is the spin stiffness.  The NLSM is scale invariant and the energy of 
the skyrmion,  $4\pi \rho_s$,
does not depend on its size or on a  global spin rotation\cite{Rajaraman-book}. By extension, the term  skyrmion is also used to name topological excitations of two component scalar field in 2D systems. In these cases  extra terms added to the NLSM make the energy, shape and size of the skyrmions depend on the details of
the Hamiltonian.
Apart from nuclear physics\cite{Multifaceted-book} skyrmions  
appear in many brands of modern physics\cite{Multifaceted-book} as two-dimensional electron gas in the quantum Hall regime\cite{Fertig:1994aa,Brey:1995aa}, liquid crystals\cite{Fukuda:2011aa}, Bose-Einstein condensates\cite{Al-Khawaja:2001aa} and  in the last years the study of skyrmions has convulsed the research of 
magnetic materials\cite{Roszler:2006aa,Muhlbauer:2009aa,Yu:2010aa,Duine:2013aa,Fert:2013aa,SampaioJ.:2013aa,Nagaosa:2013aa,Rosch:2017aa}.

In 2D ferromagnetic systems, a magnetic skyrmion describes a localized   spin texture,  
where in   going from  the core to the perimeter of the particle, the orientation of the spin full  wraps  the unit sphere.
The topology of the skyrmion is characterized by a nonzero  topological charge,
\begin{equation}
Q=\frac 1 {4 \pi} \int {\bf n} ({\bf r}) \cdot \left [\partial  _x {\bf n} ({\bf r}) \times \partial _y {\bf n} ({\bf r}) \right ] \,{\rm d} {\bf r} \, \, 
\label{TopoCharge}
\end{equation}
that indicates the number of times the unit sphere is wrapped  by the magnetization ${\bf n }({\bf r})$, when this covers
the full  real space.
The topological character of the skyrmions protects them from continuous deformation into  the uniform 
and conventional ferromagnetic state, conferring  them a long lifetime. This protection, in addition with the experimental fact that   skyrmions can be driven by low electrical current densities\cite{Jonietz:2010aa}, makes them very promising particles for spintronic devices\cite{Parkin:2008aa,Parkin:2015aa}.
Different mechanisms\cite{Nagaosa:2013aa} as long-ranged magnetic dipolar interactions\cite{Lin:1973aa},
frustrated exchange interactions\cite{Okubo:2012aa} or four-spin exchange mechanism\cite{Heinze:2011aa}, can produce stable skyrmions in magnetic systems. 
Very interesting are the skyrmions that appear in non-centrosymmetric magnets, where the competition between an antisymmetric Dzyaloshinskii-Moriya (DM) interaction\cite{Dzyaloshinsky:1958aa,Moriya:1960aa} and a symmetric exchange coupling can generate, in presence of an external magnetic field, skyrmions with size ranging from 5 to 100${\rm nm}$. The origin of the DMI  is  the lack of inversion symmetry in lattices or at the interface
of magnetic films. In ultrathin magnetic films the DMI appear as the exchange coupling between two atomic spins mediated by a heavy atom
with a large spin-orbit coupling (SOC)\cite{Fert:1980aa,Fert-1980}. 
The DMI between  classical spins can also occurs as a modified 
Ruderman-Kittel-Kasuya-Yosida \cite{Ruderman:1954aa,Kasuya:1956aa,Yosida:1957aa} interaction, where the intermediary 2D electron gas has a  Rashba
SOC\cite{Imamura:2004aa,Kim:2013aa,Banerjee:2014aa}.
Double exchange magnetic metals\cite{Gennes:1960aa}
with strong SOC may also generate DM interaction between the magnetic ions
\cite{Banerjee:2014aa,Calderon:2001aa}.

In diluted magnetic semiconductors a single electron is able to create a ferromagnetic collective ground state of a large number of  
magnetic impurities\cite{Fernandez-Rossier:2004aa,Leger:2006aa}. Also there is evidence\cite{Subramanian:1996aa,Majumdar:1998aa,Calderon:2000aa}
that a single electron can be self-trapped in a magnetic polaron.
The question then arises is if a single electron with Rashba spin-orbit coupling can be self-trapped in  a spin texture induced  in a coupled 
system of magnetic ions. By combining analytical calculations and numerical computation,  we find compelling evidence
that the   Rashba 
spin orbit coupling makes that a single electron creates and becomes self trapped in a skyrmion spin texture. 
We name this quasiparticle  as  {\it magnetic skyrmionic polaron}. 

\begin{figure}[htbp]
\includegraphics[width=8.5cm,clip]{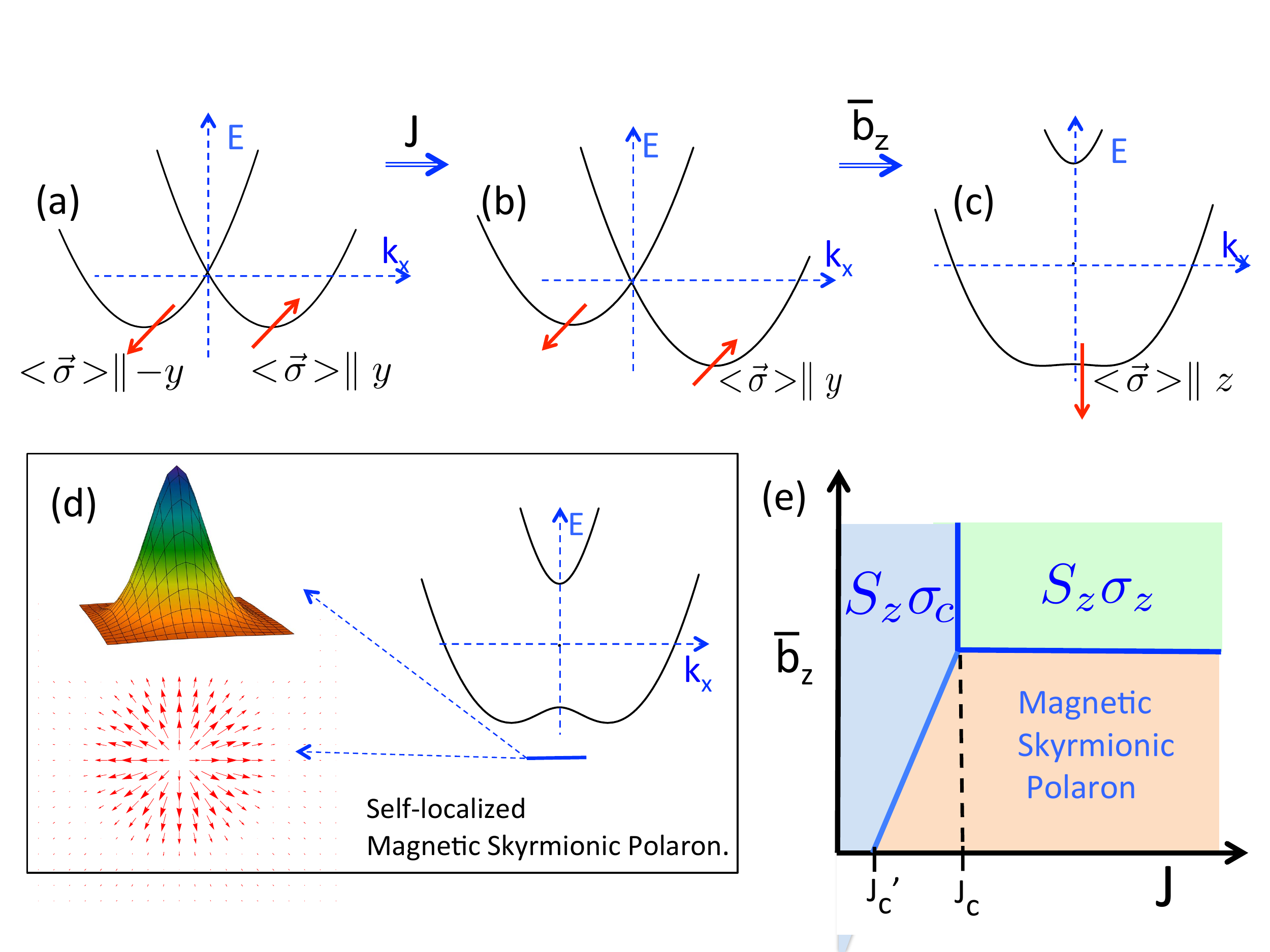}
\caption{Schematic picture of the band structure corresponding to the phases (a) $S_z \sigma _{c}$, (b)  $S_{\perp} \sigma _{\perp}$ and  (c) $S_z \sigma _z$, see text.  In (d) we indicate how an electron can be self-trapped in  a magnetic  skyrmionic polaron, with energy below the bottom of the band, and with the wave function located  on  the spin texture. In (e) we show a typical phase diagram obtained by minimizing Eq.\ref{Mini}.
 } 
\label{Figure1}
\end{figure}

\vspace{0.2cm}
\par
\noindent
The Hamiltonian for electrons moving in the conduction band   of a  2D semiconductor in the presence of a unit vector magnetization  is  
\begin{equation}
H= \frac { \hbar ^2 {\bf k} ^2} {2 m ^*} +\alpha \left ( \sigma _x k _y \! - \! \sigma _y k_x \right ) \! - \! J {\bf n}({\bf r})  \cdot  \pmb{\sigma}  \!- \! \bar  b_z  \! \int \! n_z({\bf r})  {\rm d} {\bf r}
\label{Hamil_cont}
\end{equation}
where  $\sigma _x$, $ \sigma _y$ and $\sigma _z$ are the Pauli matrices and 
$J$ is the exchange coupling constant that we consider positive. In the Hamiltonian we  also  have included a Zeeman field, $\bar b_z$,  acting on the magnetic  impurities. In Eq.\ref{Hamil_cont} $m^*$ is the effective mass and $\alpha$ the Rashba SOC.

Different trivial uniform states can surge from the coupling between the electron spin  and the magnetic impurities. For $J$=$\bar b_z$=0, the energy of the electron is $\epsilon$=$\frac {\hbar ^2}{2 m^*} (k-k_R)^2-\frac {\hbar ^2 k_R^2}{2m^*}$
with $k_R$=$\frac{m^*}{\hbar ^2} \alpha$  and  the minimum occurs at
the ring defined by  $k$=$k_R$.
Because of the SOC, the spin of a state  with momentum $k_R(\cos\beta,\sin\beta)$,  points  in the direction $< \! 
{\vec \sigma} \!>$= $(-\sin \beta, \cos\beta,0)$.  For  $\bar b_z$=0 and finite coupling between the electron spin and the magnetic ions, $J\ne0$, a single electron  selects an in-plane magnetic orientation for the system of magnetic ions. 
%
We call this phase $S_{\perp}\sigma_{\perp}$,  Figure \ref{Figure1}(b).
For $J \ne$0 and  $\bar b_z \ne 0$, ${\bf n} ({\bf r})$ polarizes along the $z$-direction and  the electron spin orientation gets a component along the $z$-direction,  canted phase $S_z\sigma_c$, until that for 
 $J$>$J_c$=$\frac {\hbar ^2 k_R^2}{2m^*}$, both the electron spin and ${\bf n} ({\bf r})$   become polarized in the $z$-direction
in a phase we call $S_z\sigma_z$,  Figure \ref{Figure1}(c). 
In this work we study the transition, as function of $\bar b_z$,  between the $S_{\perp}\sigma_{\perp}$ and the $S_z\sigma_z$ phases and show 
that the electron can be self-trapped  by   a skyrmion induced in the coupled magnetic ion system,  forming  a magnetic skyrmionic polaron, Figure \ref{Figure1}(d).

\vspace{0.3cm}
\par
\noindent
Effective spinless Hamiltonian.
Here we consider  a single electron en presence of a $Q$=1 skyrmion of the form, 
\begin{equation}
{\bf n} ({\bf r})= \left ( \cos{\theta} \sin \xi (r), \sin{\theta} \sin {\xi(r)}, \cos {\xi (r)} \right ) \, \, 
\label{texture}
\end{equation}
where $\theta$ is the polar angle  and $\xi(r)$ is a continuous function satisfying $\xi(r=0)$=$\pi$ and $\xi(r \rightarrow \infty)$=0, and we minimize the energy of the system 
withs respect the size of the skyrmion.
We use  an unitary  transformation of the Hamiltonian which makes the spin quantization direction parallel to the direction of the magnetization  ${\bf n} ({\bf r})$\cite{Tatara:1997aa,Bruno:2004aa,Finocchiaro}, i.e. $U^{\dagger} {\bf n}({\bf r}) 
\, \pmb{\sigma} U$=$\sigma _z$ with $U$=$\left ( \begin{array}{cc} u_1 ^* & - u_2 ^* \\ u_2 & u _1 \end{array} \right )$, being
$u_1$=$\cos \frac  {\xi (r)}  2 e ^{i\theta /2}$ and $u_2$=$\sin \frac {\xi (r)}2 e ^{i\theta /2}$. 
We are interested in the zero density limit in which  $J$ is the largest energy in the system and 
we neglect electrons with spin locally antiparallel to the magnetization ${\bf n} ({\bf r})$. In this approximation   we obtain the following Hamiltonian for spinless electrons,
\begin{eqnarray}
H'& = &\frac {\hbar ^2}{2 m ^*} \left ({\bf k} -\frac  e {\hbar c} {\bf a} \right ) ^2 + \frac {\hbar ^2}{ 8 m ^{* }} \left ( \frac {\partial \xi(r)}{\partial r} \right ) ^2+
\nonumber \\
&\alpha& \!\left ( \frac 1 2 \frac{\partial \xi (r)}{\partial r} \!+ i \frac {\sin{\xi (r)}}{r} \frac {\partial}{\partial \theta} \right ) \!-J- \! \bar b_z  \! \int \! n_z({\bf r})  {\rm d} {\bf r} \, ,
\label{Htrans}
\end{eqnarray} 
\begin{equation}
{\rm with} \, \, \,  \phi_0=\frac {hc} e \, \, \,  {\rm and} \, \, \, 
{\bf a} ({\bf r})= \frac {\phi _0}{4 \pi} \frac  {\cos {\xi(r)}} { r^2} (y,-x) \, \, .
\nonumber 
\end{equation}
The first two  terms in $H'$ contain the coupling between the electron and the gauge vector potential and electrostatic potential  generated by the spatial variation of the order parameter ${\bf n} ({\bf r})$, respectively. The third has its origin in the rotation of the Rashba term.  The vector potential ${\bf a}({\bf r})$ 
describes a topological magnetic field  $B_t$=$ \partial _x a _y$ -$\partial _y a_x$=$\frac {\phi _0}{4 \pi} \frac  {\sin {\xi(r)}} { r}\frac {\partial \xi(r)}{\partial r}$, that is proportional to the density of topological charge in the spin texture. 
 The skyrmion profile is well modeled\cite{Hamamoto:2015aa} as $\xi (r)$=$\pi (1-r/\lambda)$ for $r< \lambda$ and $\xi (r)$=0 for $r>\lambda$,  that describes a smooth  radial  spin texture  inside the skyrmion radius $\lambda$.
For this profile, $\frac {\partial \xi(r)}{\partial r}$=$-\pi /\lambda$, for $r$<$\lambda$ and zero elsewhere, then the second and third term of $H'$ describe a quantum dot of radius $\lambda $ and potential $\frac {\hbar ^2}{8 m^*} \frac {\pi ^2 }{\lambda ^2}-\alpha\frac {\pi }{2\lambda}$.
The average  topological  magnetic field   takes a value $\bar B_t$=$ \frac  1 {\pi}  \frac {\phi_0}{ \lambda ^2}$
that implies a magnetic length $\ell$=$\lambda / \sqrt{2}$  smaller that than the dot radius and therefore an electron located on the skyrmion should have a zero point motion energy 
$\frac {\hbar \omega _c} 2 \sim \frac {\hbar ^2} {2 m^*} \frac {2}{\lambda ^2}$. Adding all contributions the energy of the system is,
\begin{equation}
E\simeq \frac {\hbar ^2 }{2 m^*} \left ( \frac {\pi ^2}{4} +2 \right )\frac 1 {\lambda ^2} -\alpha \frac {\pi}{2 \lambda}-J+2 \pi \bar b_z \lambda ^2 \left ( \frac 1 2 -\frac 2 {\pi ^2} \right ).
\label{Mini}
\end{equation}
Minimizing this equation with respect the radius of the skyrmion, we obtain that for small values of the Zeeman field, there is always  a negative energy solution
that corresponds to the self trapping of the electron in a self-induced skyrmion spin texture. 
At ${\bar b}_z$=0 the radius of the skyrmion is $\lambda \approx \frac {3 \hbar ^2}{m^* \alpha}$, and   decreases as $\bar b_z$ increases.  

We call this quasiparticle  a  {\it magnetic skyrmionic polaron}. 
This particle competes in energy with the non-topological states $S_{\perp} \sigma _{\perp}$, $S_z\sigma_{c}$ and $S_z\sigma_z$. A typical phase diagram is shown in in Figure \ref{Figure1}(e). At large values of $\bar b_z$ and $J$ all spins polarize in $z$-direction and the ground state is $S_z \sigma _z$, for large values of $\bar b _z$ but moderate values of $J$,  the magnetic ions polarize in the $z$-direction, but the electron spin is canted, phase 
$S_z \sigma _{c}$. For large values of $J$ and moderate values of $\bar b_  z$ the electron self-traps in a magnetic skyrmionic polaron.  At $\bar b _z$=0,  the transition between  the $S_z\sigma _{c}$ and the magnetic polaronic phase occurs at $J_c'$. 
The difference between  $J_c' $-$J_c$ is  the binding energy of the magnetic skyrmionic polaron at $\bar b_z$=0.

\begin{figure*}[htbp]
\includegraphics[width=17cm,clip]{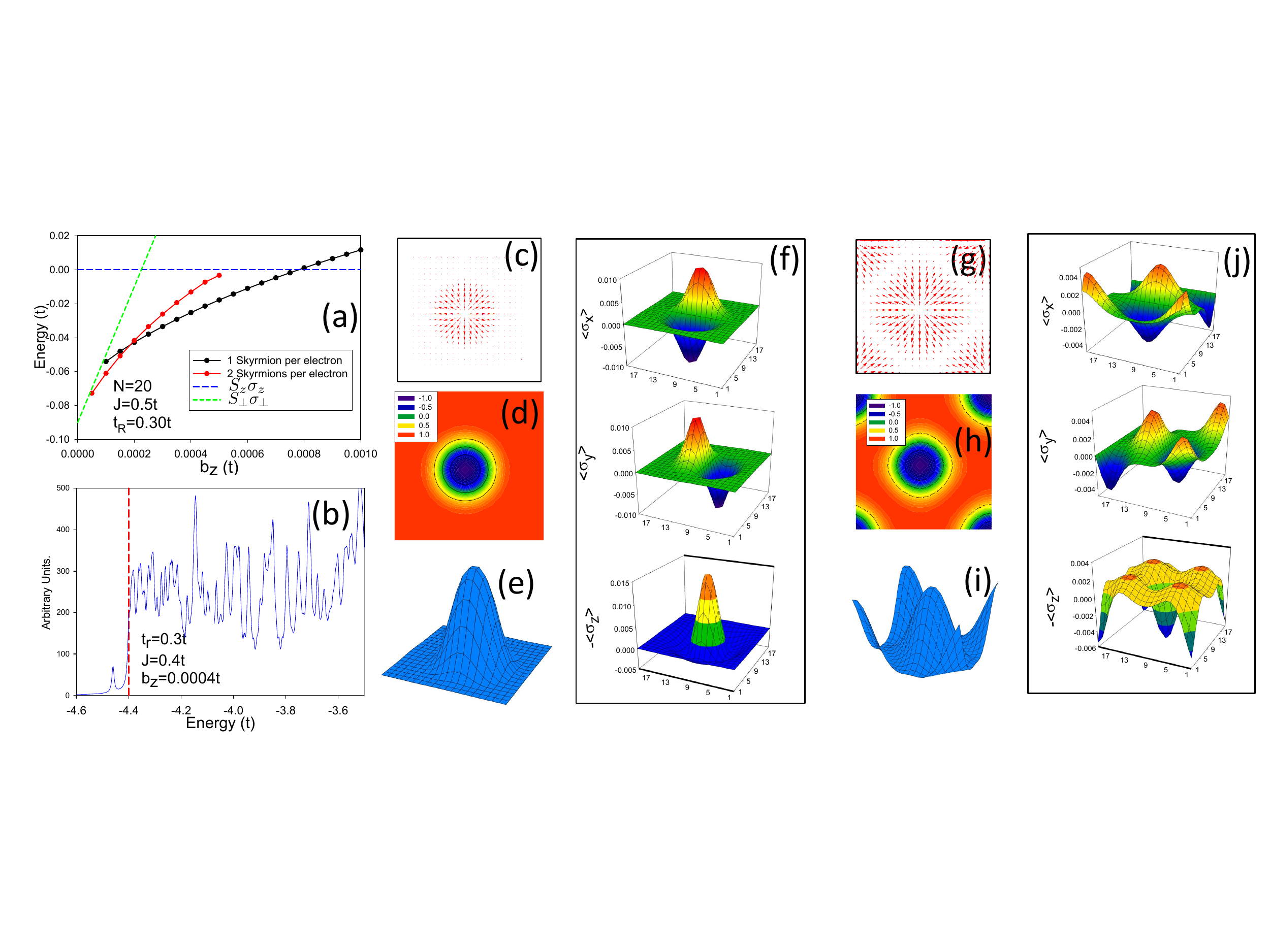}
\caption{(a) Energy of different states, as function of $b_z$, for $J$=0.5$t$, $t_R$=0.3$t$, and $N=20$, as obtained from tight-binding calculations.. The energies are plotted with respect the energy of the $S_z \sigma _z$ state. (b)
Electron density of states for the case $b_z$=0.0005$t$,  $J$=0.4$t$, $t_R$=0.3$t$, and $N=50$. The vertical red line indicated the position of the bottom of the electronic band. The sharp peak below the band correspond to the magnetic skyrmionic polaron. (c) and (d) show the spin texture corresponding to   a $Q$=1 skyrmion at $b_z$=0.0005$t$, the other parameters the same
than in (a). In (e) we plot the topological charge density for the same parameters than in (c).  In (g)-(j) we plot the same magnitudes than in (c)-(f)  for the case $Q$=2 and $b_z$=0.0002$t$.
} 
\label{Figure2}
\end{figure*}

\vspace{0.1cm}
\par
\noindent
 The results obtained from  the effective spinless Hamiltonian  Eq.\ref{Htrans}, provide a 
physical insight of how an electron can be self-trapped  in a skyrmion.  In the following we perform tight-binding calculations in order to check the 
validity of the previous conclusions and also for studying  a wider range  of values of the exchange coupling  $J$ and obtain a more accurate  description of the magnetic skyrmionic polarons. We consider a square lattice, 
first-neighbors hopping tight-binding Hamiltonian\cite{Hankiewicz:2004aa},
\begin{eqnarray}
&H&   = \! -t \! \! \sum_{i,j,\sigma} \! c^+_{i,\sigma} c_{j,\sigma} +
\! t_R  \!  {\Bigl ( } \sum_j \! \! \! -i (c^+_{j,\uparrow} c_{j+{a_ y},\downarrow} +c^+_{j,\downarrow} c_{j+{a_ y},\uparrow}  ) + \, \, \, \, \, \, \nonumber \\  &(& \! \! c^+_{j,\uparrow} c_{j+a_x,\downarrow} \! - \! c^+_{j,\downarrow} c_{j+a_x,\uparrow}) \! + \!h.c.\Bigr )
\! -\! J \!\sum_{i} {\bf S}_i \, {{\pmb \sigma}}_i \!  \!-\! b_z \!\! \sum _i S_{i,z}\, 
\label{HTB}
\end{eqnarray}
where $c^+_{i,\sigma}$ creates an electron at site $i$ with spin $\sigma$,
$t$=$\hbar ^2/2m^*a^2 $, $t_R$=$\alpha/2a$,  ${\pmb \sigma_i}$=$c^+ _{i,\alpha} 
{\pmb \sigma} _{\alpha,\beta} c_{i,\beta}$ is the electron spin operator, ${\bf S}_i$ is the impurity spin normalized to unity  and  $\vec \delta$=$(\pm a \hat x,\pm a \hat y)$ ,being $a$ the lattice 
spacing.
We consider a single electron  in  a supercell containing  $N\times N$ sites and use periodic boundary conditions,  so that the density of charge in the system
is $\frac 1 {N^2 a^2}$. 
The Hamiltonian Eq.\ref{HTB} is solved self consistently  in a process where after diagonalizing and obtaining the  wave-function of the lowest energy  state in presence of a spin texture, we recalculated the spin texture created by the expectation value of  the electron spin operator and $b_z$  and then repeat the process until  input and output coincide.
Solving Eq.\ref{HTB}, we obtain the  electron energy and wave-function, as well that  the orientation of the magnetic ions  $\{ {\bf S}_i \}$.  For a given spin configuration the topological charge is obtained by following the prescription of Berg and Lusher\cite{Berg:1981aa}. 
We divide each basic unit square of the lattice into two triangles.  
The three spins at the corners of  each triangle $l$   define a signed area on the unit sphere $A_l$, and the topological charge in the supecell is given by 
$Q$=$1/4\pi \sum _l A_l$, where the sum runs over all triangles in the system.

By starting the self-consistent procedure with seeds corresponding to different spin textures, we obtain solutions with different  topological charge in the unit cell. In the tight-binding calculation we do not impose any constrain to the size and shape of the skyrmions.  In Figure \ref{Figure2}(a)  we plot the energy per unit cell of different states for the case $t_R$=0.3$t$, $J$=0.5$t$ and $N$=20, as function of the Zeeman field $b_z$. For these large values of $J$, the relevant uniform non-topological phases are  $S_z \sigma_z$, with energy   -4$t$-$J$-$N^2b_z$ that is the ground state for  large values of $b_z$ and  $S_{\perp} \sigma _{\perp}$ with energy
 -4$t$-$t_R^2/t$-$J$ that is the relevant phase for   $N^2b_z < t_R^2/t$.
We see in Figure \ref{Figure2}(a) that there is a range of Zeeman fields for which the low energy state corresponds to a $Q$=1  magnetic skyrmionic polaron. This is evident in Figure \ref{Figure2}(c)-(d) where we plot the $x-y$ and the $z$-component of the ion spins for $J$=0.5$t$, $t_R$=0.3$t$ and $b_z$=0.0005$t$. Clearly, there is an isolated skyrmion located at the center of the unit cell. The coupling between the electron and the magnetic ions is also reflected in the electron spin density shown in 
Figure \ref{Figure2}(f), that shows the same symmetry and shape than the ions spin texture.
The self-trapping of the electron by the skyrmion and the binding energy is also reflected in the electron density of states, Figure \ref{Figure2}(b); below the bottom of the electron band, -4$t$-$J$, there appears a sharp peak containing exactly one electron and that corresponds to the magnetic skyrmionic polaron.  
In Figure \ref{Figure2}(a) we notice that at very low Zeeman fields, the low energy phase corresponds to a topological charge $Q$=2  in the unit cell. This phase occurs because for small values of $b_z$, the size of the skyrmion increases and skyrmions centered in neighbor unit cells overlap and interact, being energetically favorable for the system to creates  a skyrmion crystal with two $Q$=1 skyrmions per unit cell, Figure \ref{Figure2}(g)-(i), and a single electron per unit cell Figure \ref{Figure2}(j).

\begin{figure}[htbp]
\includegraphics[width=8.5cm,clip]{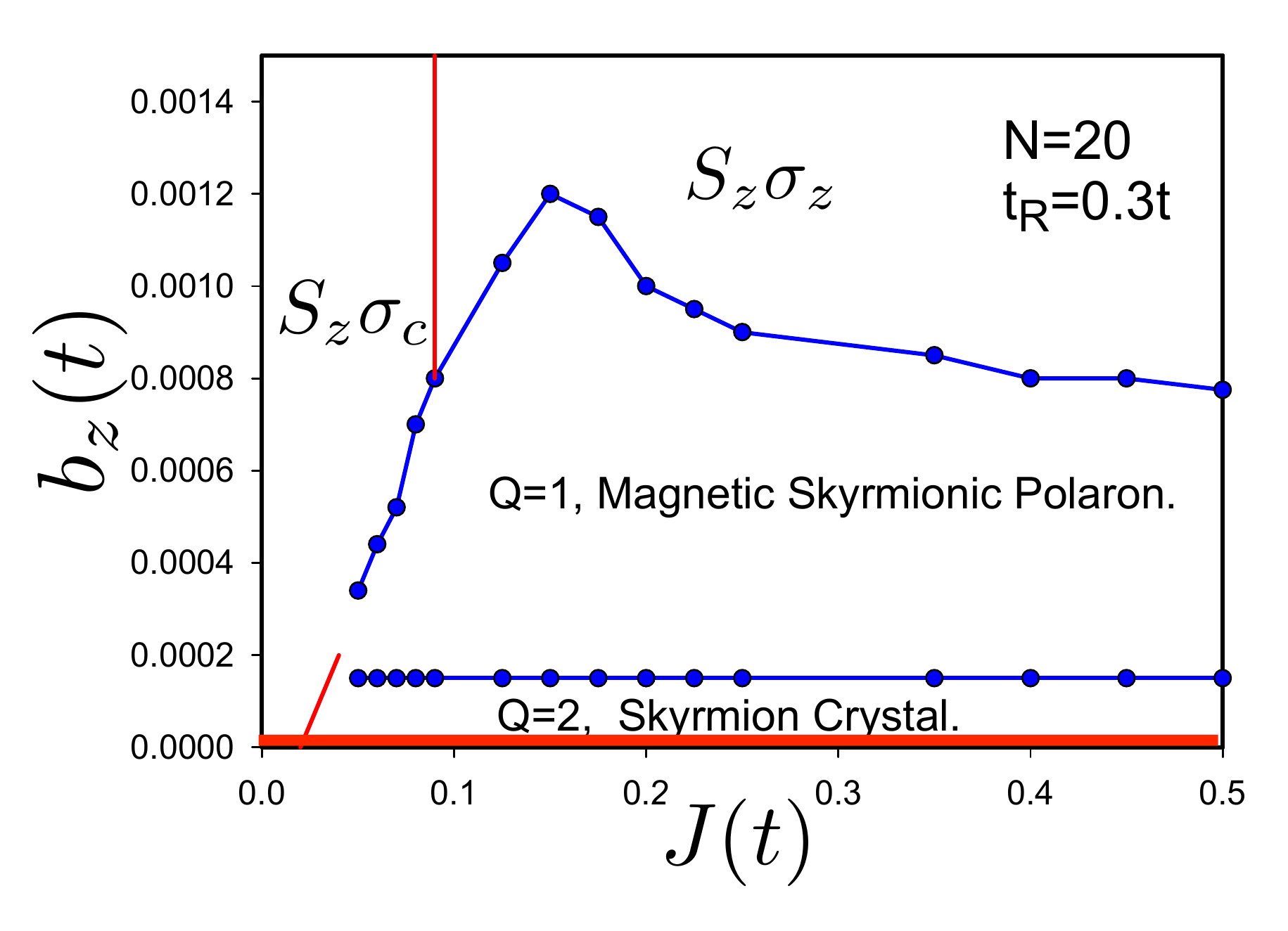}
\caption{Phase diagram, $b_z$ versus $J$, for $t_R$=0.3t and calculated in a supercell with $N$=20. The phase $S_{\perp}\sigma _{\perp}$ phase corresponds to the thick red  line, and it is the ground state at 
very small values of $N^2 b_z$.
 } 
\label{Figure3}
\end{figure}

In Figure \ref{Figure3} we show a phase diagram $b_z$ versus $J$,  for $t_R$=0.3t obtained in a supercell with $N$=20.  The main differences when comparing with the obtained 
in the effective spinless Hamiltonian, Figure \ref{Figure1}(e) are,  first there is a small but finite region of values of $b_z$ where the ground state is
$S_{\perp} \sigma_{\perp}$ and second, by increasing $b_z$ in the transition from the $S_{\perp} \sigma_{\perp}$ state to the magnetic skyrmionic polaron phase, there appears a region where there are two skyrmions per electron,  crystallizing in a square lattice. Both effects occur because of the small size of the  unit cell, the energy associated with the polarization of the magnetic ions in the $z$-direction is $b_z N^2$, and therefore as larger is $N$,  smaller is the region  in the $J$-$b_z$ parameter space where $S_{\perp} \sigma _{\perp}$ is the ground state. In the same way, as $N$ increases the overlap between magnetic skyrmionic polaron decreases and the Q=2 region disappears in the phase diagram.
\vspace{0.2cm}
\par
\noindent
{Dynamics of magnetic skyrmionic polarons. The dynamics of the impurity spins, treated as classicals,  is dictated by the Landau-Lifshitz equation\cite{Slonczewski:1996aa},
\begin{equation}
\partial _t {\bf S}_i =\frac 1 {\hbar}  {\bf H _i} \times {\bf S}_i +\alpha_G {\bf S}_i \times \partial _t {\bf S}_i \, \, 
\label{LL}
\end{equation}
where $\alpha_{G}$<<1 is the phenomenological Gilbert damping constant and ${\bf H}_i  $ is the local   effective field, written in energy units, that is derived form te Hamiltonian,
${\bf H}_i =-\frac {\partial H} {\partial {{\bf S}_i}}$.
In the previous equation we have not included a current-induced torque because although the spin texture is charged it is placed in an insulator materials\cite{Knoester:2014aa,Hals:2014aa}. Equation \ref{LL} describes both the deformation and the dynamics of the skyrmions. The excitation  of the internal modes  of the skyrmions have a finite frequency, that increases with the presence of  an external magnetic filed and this justify  the treatment  of skyrmions as rigid particles\cite{Lin:2013aa}, that move across the sample without distortion.
Therefore, the dynamics of skyrmions is defined by the position $(X , Y)$
and velocity $(V_x,V_y)$ of its center of mass.
The motion of the center of mass is obtained by  integrating the Landau-Lifshitz equations following the method developed by  Thiele\cite{Thiele:1973aa,Tretiakov:2008aa,Diaz:2016aa}, and for a magnetic texture as that described in Eq.\ref{texture} we obtain,
\begin{eqnarray}
F_x & = & - G M V_y + \alpha_GM D V_x \nonumber \\
F_y & = & G M V_x+\alpha _G M D V_y \, ,
\end{eqnarray}
here $M$ is the magnetic moment per unit area in the magnetic system, $G$=$4\pi Q$,
$D$=$ \pi \int _0 ^{\infty} d{\bf r} { \left ( \frac {\sin  ^2 \xi(r) } {r} + r \left ( \frac {\partial \xi (r)}{\partial r } \right )^2 \right ) }$
 and ${\bf F}$ is the so called generalized force  acting on the skyrmion due to impurities, magnetic fields or boundary conditions\cite{Nagaosa:2013aa}. 
Magnetic skyrmionic polarons are charged particles and therefore they also response to electromagnetic fields. In the case of an electric field $E$=$(E_x,0)$, the skyrmion  acquires a velocity,
\begin{eqnarray}
V_x&=&  \frac {\alpha MD}{(\alpha _G M D)^2 + M^2 G^2} e E_x \\
V_y&=&-  \frac { MG}{(\alpha _G  M D)^2 + M^2 G^2} e E_x \, .
\end{eqnarray}
The magnetic skyrmionic polaron  steady velocity in the direction of the electric field is determined by the balance between electric field and damping. In the limit of vanishing damping the particle shows a Hall effect with $V_x$=$0$ and $V_y$=$\mu E_x$, that defines a mobility $\mu$=$-e/MG$, that for a density of magnetic moments of the order of $\hbar/ {\rm nm} ^ {-1}$ takes values  $\mu\sim 1\times 10^{-4} {\rm m}^2 /{\rm V s}$ that implies drift velocities $v_H \sim0.01{\rm m/s}$ for applied electric fields $E_x$=$10^{2}$eV/m, comparable or larger  to  that of magnetic skyrmions driven by electric current\cite{Schulz:2012aa,Jiang:2017aa}. Large drift velocities have been also predict for skyrmions placed on top of topological insulators\cite{Hurst:2015aa}.

\begin{figure}[htbp]
\includegraphics[width=8.5cm,clip]{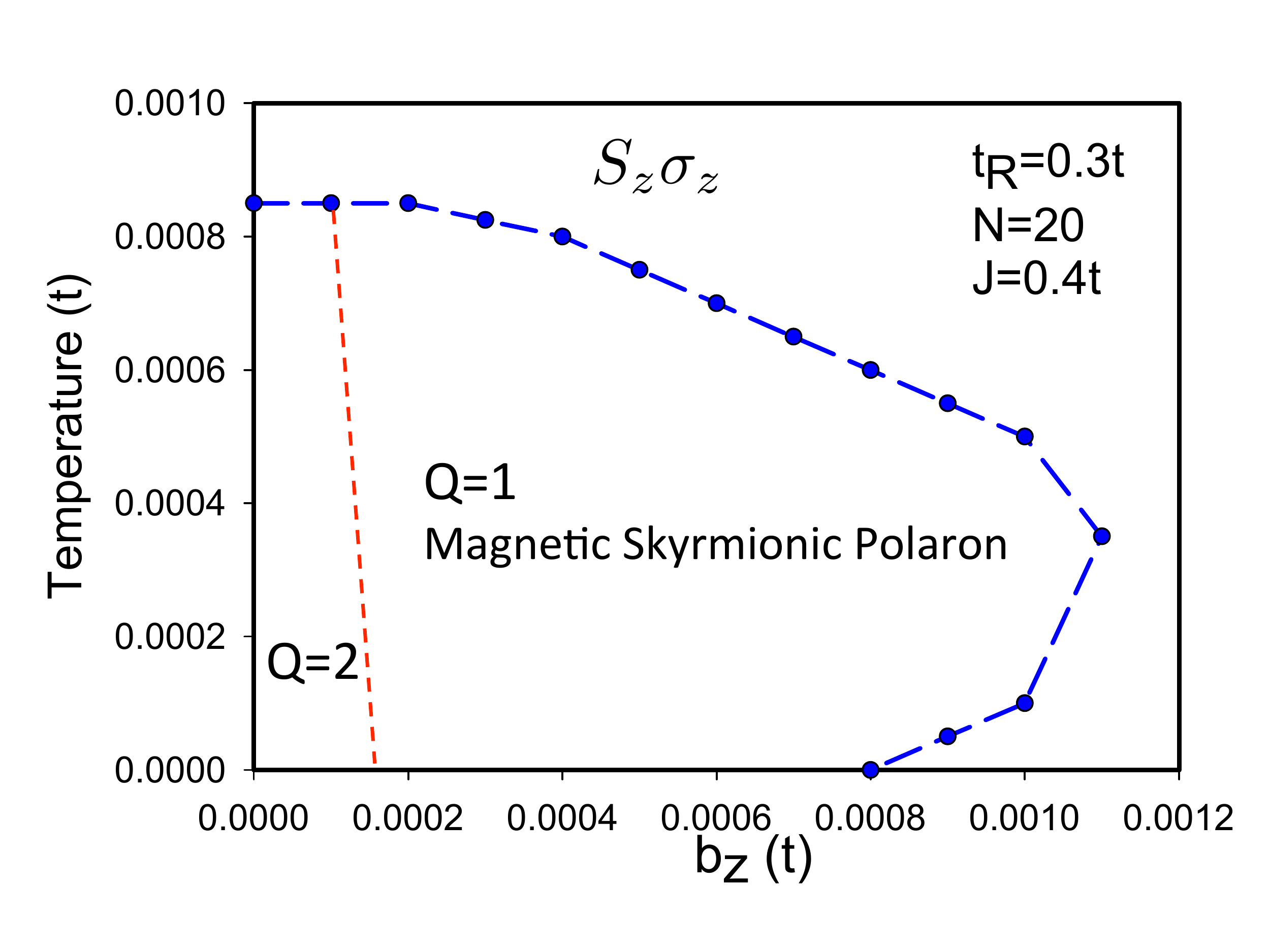}
\caption{Temperature-$b_z$, phase diagram, for $t_R$=0.3t,  $J$=0.4t  and calculated in a supercell with $N$=20.
 } 
\label{Figure4}
\end{figure}

\vspace{0.2cm}
\par
\noindent
We estimate the effect of the temperature on the existence of magnetic skyrmionic polarons  in the mean field approximation. The effective magnetic filed acting on the normalized spin impurity ${\bf S}_i$ is,
\begin{equation}
{\bf B}_i =(- J < \! \sigma _{x,i} \! >,- J < \! \sigma _{y,i} \! >,- J < \! \sigma _{z,i} \! >-b_z) \, , 
\end{equation}
where $< \! {\pmb \sigma} _i \! >$ us the expectation value of the electron spin operator at site $i$. In the mean field approximation  the free energy of classical spin $S_i$ is,
\begin{equation}
{\cal F}_i = -k_B T  \ln { \left [ 2 \frac { \sinh (B_i/k_BT)}{B_i /k_B T} \right]} \, 
\end{equation}
from which the expectation value of the spin impurity is\cite{Brey:2006aa}
\begin{equation}
< \! {\bf S}_i \!> =< \! S_i \! > \! \frac {{\bf B}_i}{B_i}\, \, {\rm with} \, \, <\! S_i \!> = \! \frac 1 {\tanh(B_i/k_B T)}\!  - \! \frac {k_B T} {B_i}.
\label{self-tem}
\end{equation} 
By solving self-consisting Eq.\ref{HTB}, with the values of the ion spin ${\bf S}_i$ replaced by $< \! {\bf S}_i \!>$ , Eq.\ref{self-tem}, we obtain the temperature phase diagram for the occurrence of magnetic skyrmionic polarons.  In this approximation we assume that the magnetic skyrmionic polaron binding energy os larger than the temperature and the only participating electron state is that with lowest energy. 
The main effect of temperature is  the reduction of the expectation value of the classical spins, that results in an effective reduction of the Rashba spin orbit coupling in Eq.\ref{HTB}. 
In Figure \ref{Figure4} we plot the phase diagram temperature-$b_z$ for the parameters $t_R$=0.3$t$, $J$=0.4$t$ and a unit cell $N=20$. When increasing temperature the reduction of the effective Rashba coupling induces a  transition from the skyrmionic phase  to the $S_z\sigma _z$ phase. The reentrance of the $S_z \sigma _z$ phase at large values of $b_z$ reflects the maxima occuring in the $b_z$-$J$ phase diagram, Eq.\ref{Figure3}.

\vspace{0.2cm}
\par
\noindent
In closing, in this paper we have shown, using analytical estimations and numerical calculations, that a single electron moving
in a band with a strong Rashba spin-orbit coupling and coupled to magnetic ions  can be self-trapped by forming a skyrmion spin texture.  We have calculated the phase diagram for the existence of this
magnetic skyrmionic polaron, as function of the exchange coupling, Rashba SOC, temperature and external magnetic field. Also, we have analyzed the dynamics of the skyrmionic polaron and found that in presence of an electric field, the skyrmion will show a Hall effect with a rather large mobility and drift velocity. 

In this work we have studied a  2D semiconductor  in the  limit of  very low electron density. For higher electron densities, a well-defined RKKY interaction between magnetic impurities should develop.  In the presence of Rashba spin-orbit coupling, this interaction should induce effective exchange and DM interactions between the ion spins. In this high-density regime, a phase diagram with a helical, ferromagnetic and skyrmion crystal phases, as that presented in ref.\cite{Banerjee:2014aa}, is expected. In this high density regime, the skyrmions in the crystal phase should overlap strongly and the resulting electronic density should be slightly modulated with a small charge accumulation near the skyrmion core\cite{Freimuth:2013aa}.

The existence of magnetic skyrmionic polaron  requires a semiconductor with a large Rashba SOC and a large exchange coupling between the electron spin and the magnetic ions. Recently it has been 
reported soft X-ray ARPES on  epitaxially grown thin films of Ge$_{1-x}$Mn$_x$Te\cite{Krempasky:2016aa}. For a concentration $x$=5.4{\%} of Mn,  the measured  band structure parameters are,
$m^*$=0.2, $\lambda$=3eV\AA, and $J$=0.11eV, that correspond to the following tight-binding parameters, $t$=0.53eV, $t_R$=0.25eV and J=0.06eV obtained using a lattice constant $a \sim 6$\AA\cite{Knoff:2015aa}. 
From these  numbers, and comparing with the results presented in Figure \ref{Figure2} and Figure \ref{Figure3} we conclude that it is possible that magnetic skyrmionic  polarons appear in Ge$_{1-x}$Mn$_x$Te at very low densities and  temperatures lower than 10K.
For a concentration of Mn ions of 5{\%} the magnetic skyrmionic polarons will be present up to rather large magnetic fields, $>$20T.
Recently obtained quasi two-dimensional ferromagnetic materials\cite{Samarth:2017aa,Gong:2017aa,Huang:2017aa}, are expected to have a strong Rashba spin orbit coupling and therefore are also good candidates to host magnetic skyrmionic polarons.

\setlength{\parskip}{0.1cm}
\par
\noindent

\begin{acknowledgement}
Author acknowledges  H.A.Fertig for helpful discussions. This work has been partially supported by  the Spanish
MINECO Grant No. FIS2015--64654-P.
\end{acknowledgement}
\providecommand*\mcitethebibliography{\thebibliography}
\csname @ifundefined\endcsname{endmcitethebibliography}
  {\let\endmcitethebibliography\endthebibliography}{}


\end{document}